\documentclass[prb,twocolumn,showpacs,superscriptaddress]{revtex4}
\usepackage{mathptmx}
\usepackage[T1]{fontenc}
\usepackage[latin9]{inputenc}
\usepackage{verbatim}
\usepackage{amsmath}
\usepackage{graphicx}
\usepackage[usenames,dvipsnames]{color}

\newcommand{\bq}{\begin{mathletters}}
\newcommand{\eq}{\end{mathletters}}
\newcommand{\beq}{\begin{eqnarray}}
\newcommand{\eeq}{\end{eqnarray}}
\newcommand{\beqq}{\begin{eqnarray*}}
\newcommand{\eeqq}{\end{eqnarray*}}

\newcommand{\rmi}{{\rm i}}

\newcommand{\hh}{{\bf H}}
\newcommand{\ee}{{\bf E}}
\newcommand{\ww}{{\bf W}}

\newcommand{\rr}{{\bf r}}
\newcommand{\eee}{{\bf e}}

\newcommand{\andrey}[1]{ \textcolor{red}{({\bf Andrey:} #1)}}
\newcommand{\sergei}[1]{ \textcolor{blue}{({\bf Sergei:} #1)}}

\begin{document}

\title{Asymmetric transmission in planar chiral split-ring metamaterials: \\
Microscopic Lorentz-theory approach}
%\title{Lorentz-theory homogenization of planar chiral split-ring metamaterials}

\author{Andrey V. Novitsky}
\email{anov@fotonik.dtu.dk}
\affiliation{DTU Fotonik, Department of Photonics Engineering, Technical University of Denmark, {\O}rsteds Plads 343, DK-2800 Kgs. Lyngby, Denmark}
\affiliation{Department of Theoretical Physics, Belarusian
State University, Nezavisimosti Avenue 4, 220030 Minsk, Belarus}

\author{Vladimir M. Galynsky}
\affiliation{Department of Theoretical Physics, Belarusian
State University, Nezavisimosti Avenue 4, 220030 Minsk, Belarus}

\author{Sergei V. Zhukovsky}
\email{szhukov@physics.utoronto.ca}
\affiliation{DTU Fotonik, Department of Photonics Engineering, Technical University of Denmark, {\O}rsteds Plads 343, DK-2800 Kgs. Lyngby, Denmark}
\affiliation{Department of Physics and Institute for Optical Sciences, University of Toronto, 60 St.~George Street, Toronto, Ontario M5S 1A7, Canada}
\affiliation{Theoretical Nano-Photonics, Institute of High-Frequency and Communication
Technology, Faculty of 
%Electrical, Information and Media 
Engineering,
University of Wuppertal,
Rainer-Gruenter-Str.~21, D-42119 Wuppertal, Germany}

\begin{abstract}
The electronic Lorentz theory is employed to explain the optical properties of planar split-ring metamaterials. Starting from the dynamics of individual free carriers, the electromagnetic response of an individual split-ring meta-atom is determined, and the effective permittivity tensor of the metamaterial is calculated for normal incidence of light. Whenever the split ring lacks in-plane mirror symmetry, the corresponding permittivity tensor has a crystallographic structure of an elliptically dichroic medium, and the metamaterial exhibits optical properties of planar chiral structures. Its transmission spectra are different for right-handed vs.~left-handed circular polarization of the incident wave, so the structure changes its transmittance when the direction of incidence is reversed. The magnitude of this change is shown to be related to the geometric parameters of the split ring. The proposed approach can be generalized to a wide variety of metal-dielectric metamaterial geometries.
\end{abstract}


%%% ======================================================================

\maketitle

\section{Introduction\label{sec:1}}

The concept of metamaterials have drawn a strong interest
ever since fabrication and characterization of such structures became
feasible. The ability to engineer the metamaterial elements (``meta-atoms'')
in a largely arbitrary fashion adds a whole new dimension of freedom
in material science. Artificial materials show promise for a wide
range of unusual physical phenomena, some of which are rare
or absent in nature. Examples include negative refraction, \cite{ShalaevNeg}
hyperbolic dispersion relation resulting in anomalously high density of states in a wide frequency range, \cite{NarimanovHyp} or support for transformation optics, \cite{UlfTrans,PendryTrans,ShalaevTrans}
offering superior degree of control over light propagation and
guiding.

\begin{figure}[b]
\includegraphics[width=1 \columnwidth]{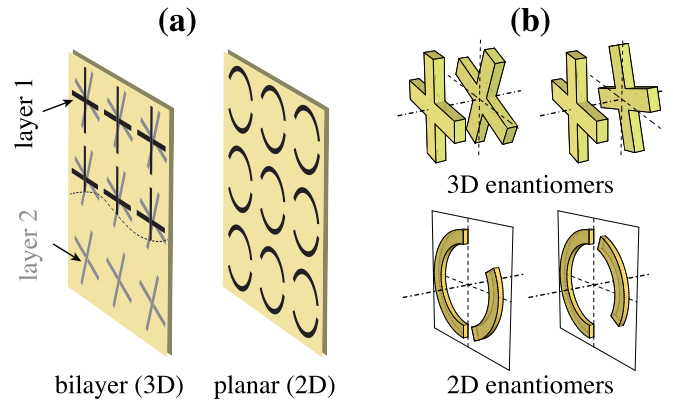}
\caption{(Color online) (a) Schematic of a bilayer (3D) vs.~planar (2D) chiral metamaterial;\textbf{
}(b) illustration of 3D and 2D enantiomers \label{fig:PCMs}}
\end{figure}

More specifically, if the shape of the meta-atoms is chiral, i.e., when the meta-atom cannot be superimposed with its own mirror image, such metamaterials resemble naturally occurring optically active media and outperform them by orders of magnitude. \cite{GiantGyro,GiantGyroZhel,CarstenGyro,WegenerGyro} More recently, following a seminal paper by Plum et al, \cite{ZhelPRL} \emph{planar chiral metamaterials} (PCMs) were introduced. They consist of planar meta-atoms on a flat substrate \cite{ZhelPRL,DrezetPCM-OE,PlumCSR} that cannot be superimposed with their {\em in-plane} mirror image without being lifted off the plane. In other words, truly chiral metamaterials possess distinct 3D enantiomers while PCMs do not, but can be said to possess distinct 2D enantiomers [see Fig.~\ref{fig:PCMs}(b)].

A notable feature of the PCMs is their asymmetry in propagation of
electromagnetic waves incident from opposite directions or having
different (right-~or~left-handed) circular polarization. \cite{ZhelPRL}
This directional asymmetry results from the fact that the wave propagating
in the opposite direction effectively interacts with the mirror image
of the original structure. Therefore it is a manifestation of asymmetry
between 2D enantiomers. This makes PCMs distinct from
3D chiral metamaterials \cite{Plum3D,WegenerGyro,CarstenGyro} where 3D enantiomers
do not flip when the wave propagation direction is reversed.

Directional asymmetry is also known to exist in other environments
such as the Faraday cell. In this case, however, it is usually attributed to non-reciprocity due to the presence of magnetic field, while PCMs do not have such non-reciprocity.
Further analysis reveals that both in Faraday media and in
optically active liquids the waves whose polarization state does not change
during propagation (polarization eigenstates) are circularly
polarized and \emph{counter-rotating}, i.e., they come in pairs of
eigenwaves one of which has right-handed (RH) and the other left-handed
(LH) circular polarization. In contrast, PCMs have polarization eigenstates that are elliptical
and \emph{co-rotating}, \cite{ZhelPRL} meaning that their handedness
is the same for both eigenwaves in the pair.

Thus, while 3D chiral metamaterials are simply an analogy of bi-isotropic (e.g., optically active liquids) or bianisotropic (gyrotropic) media, PCMs clearly represent a distinct type of electromagnetic materials, apparently without an immediate naturally occurring counterpart. Hence, explaining the exotic optical properties of PCMs certainly poses
an exciting problem in theoretical electrodynamics. On an abstract
crystallographic level, it has been shown recently that
a combination of birefringence and circular dichroism provides a basis
for optical manifestation of planar chirality, and that an elliptically
dichroic medium should exhibit characteristic optical
properties of a PCM. \cite{ourOL}

On a more involved, microscopic level, several explanations of PCM operation have been developed. Most works attribute the dichroism to polarization-sensitive excitation of dark plasmonic resonant modes in meta-atoms. In a variety of PCM designs, this polarization sensitivity is attributed to magnetic dipole excitations that become ``trapped'' due to poor coupling with the incident wave. \cite{PlumCSR,Zhel09Collapse,ZhelTHz}
Other accounts \cite{Falk09} build up on earlier multipole treatment
of metamaterials \cite{Falk08} to express the meta-atom's dichroic
response based on collective dipole oscillations in its individual segments. Still other works attempt to formulate the theory of PCMs by treating the meta-atom as a plasmonic oligomer and calculating its response in coupled dipole approach.\cite{ourFiO10,ourAPB11}

However, an {\em ab initio} theoretical description of PCMs, which would explain their dichroic behavior on a microscopic level and independently of a particular PCM geometry, is still missing. A first attempt, based on considering the response of the meta-atom in terms of the dynamics of individual electrons (Lorentz theory), was made in Ref.~\onlinecite{ourOL}. That attempt, however, was only successful in arriving at the right form of the metamaterial's effective dielectric permittivity tensor, reproducing qualitatively correct optical properties of a PCM. A more rigorous, quantitative description with a detailed analysis of the applicability range for the Lorentz theory is still called for.

In this paper, we make a further step towards an {\em ab initio} description of metamaterials and present a detailed route to arrive at the chiral properties for the asymmetric split-ring PCMs.\cite{PlumCSR} We start by considering the dynamics of free electrons in an arc-shaped metallic element in presence of an external electromagnetic wave. The electron motion is shown to be determined by (i) screening forces caused by the charge redistribution in the external field, and (ii)  electromotive forces caused by currents flowing in the segment as the charges redistribute. These forces exert spring-like and inertia-like effects on an electron, respectively, and can be associated with capacitance-like and inductance-like contribution in an equivalent LC-circuit. Together, these two kinds of forces give rise to a particle plasmon resonance. Its properties are derived directly from geometric dimensions of the arc segment and material properties of the metal without the need for phenomenological parameters. By considering two such segments in close proximity, the electromagnetic response of a split-ring meta-atom is evaluated.

In order to connect the meta-atom's response to the exotic properties of PCMs described previously, a standard homogenization procedure \cite{Ishimaru,Vinogradov} is then employed to derive the effective dielectric permittivity tensor for the PCM. Since the PCM in question is a surface rather than a bulk material,\cite{Baena,Alu} this procedure should not be regarded as a true homogenization, and a multipole-expansion approach should give a more accurate physical description. \cite{Falk08,Falk09,Yatsenko,Falcone} Still, interpreting a metafilm as an effective medium can be feasible,\cite{Holloway} and the resulting effective permittivity tensor is shown to be valid for normal incidence of light in the frequency range of the fundamental resonance of the split ring. Moreover, it turns out to have the crystallographic structure of elliptically dichroic media, as would be expected for planar chiral materials. \cite{ourOL} A similar approach was recently applied to calculate the THz field enhancement in a nanoslit,\cite{ourTHz}  the results showing a good agreement with experiment.\cite{Seo}

The optical properties (transmission and reflection spectra) of the PCM are then calculated using  wave operator based extension of standard transfer-matrix method. \cite{BBL,BorzdovTheory} In agreement with previous theoretical and experimental results, \cite{ZhelPRL,PlumCSR} the spectra are sensitive to whether the incident wave has left- or right-handed circular polarization. This difference, which translates to directional asymmetry, \cite{PlumCSR} can be used to quantify the strength of planar chiral properties. The dependence of this strength on the geometrical parameters of the split ring is analyzed. When the split ring is symmetric, i.e., when there are no distinguishable 2D enantiomers, optical manifestation of planar chirality is seen to vanish as required by symmetry and reciprocity constraints. \cite{Tretyakov} Maximum chiral properties are observed when the enantiomers are most distinct.
% and when direct coupling between segments in the split ring is reduced.

%Based on the proposed, the design principles of model easily allows to analyze the dependence of planar chiral properties on the split ring design parameters, which can be used to formulate optimization principles for such PCMs.
The proposed approach can be relatively straightforwardly extended to a wider variety of shapes for planar meta-atoms. Moreover, the model lends itself to an extension along the lines of Refs.~\onlinecite{Falk08,Falk09}  to the cases of oblique incidence and non-planar meta-atom shapes. The results obtained can also be generalized from a single PCM to PCM-based multilayers, inasmuch as such generalization can be performed.\cite{Andriyeuski}

The rest of the paper is structured as follows. In Section \ref{sec:2}, the Lorentz theory is employed to arrive at the electromagnetic response of a single split-ring meta-atom. Section \ref{sec:3} follows with the procedure to arrive at the effective permittivity tensor of the PCM. The structural properties of this tensor are analyzed, and calculation of the PCM's optical spectra is outlined. In Section \ref{sec:comp}, the results on calculated spectral properties of split-ring PCMs are presented and compared with numerical simulations. In Section \ref{sec:geom}, the relations between chiral properties and meta-atom geometry are systematically analyzed. Finally, Section \ref{sec:5} summarizes the paper and outlines future extensions for the proposed theory.

%%%=================================================================

\section{Response of a split-ring meta-atom\label{sec:2}}

As an example of a planar chiral metamaterial, we consider a two-dimensional array of chiral split rings (CSRs). The corresponding meta-atom is shown in Fig.~\ref{fig:1}. It comprises a metallic ring broken into two segments in an asymmetric fashion, so that two 2D enantiomers can be distinguished [see Fig.~\ref{fig:PCMs}(b)]. CSR metamaterial was chosen for its relative geometrical simplicity and for availability of previous experimental results. \cite{PlumCSR}

The lateral width of the ring $d$ is assumed to be much smaller than its radius $R$, and its thickness  $h$ is even smaller.
%, comparable with the skin depth of the metal.
The ring sits atop a thicker dielectric substrate.
%, so the whole structure can be imagined as consisting of a thin "metamaterial layer" stacked with the substrate layer.
The metal of the ring is taken to be copper with complex dielectric permittivity $\varepsilon_m$,\cite{CopperParms} and the permittivity of the substrate is $\varepsilon_d$. Both materials are non-magnetic. A monochromatic incident wave with electric field $\ee = \ee_0 \exp(-\rmi \omega t)$ is assumed to illuminate the material, $\omega$ being the angular frequency of the wave.
%We seek to obtain the effective parameters of the "metamaterial layer" and use them to arrive at the spectral and polarization properties of the whole structure.

\subsection{Dynamics of the electrons}

We begin by considering the motion of an electron in a finite-sized metallic inclusion. Each electron is affected by the driving force $e \ee$ originating from the external electric field, the friction  force $-m \gamma {\bf v}$ due to losses in the metal, the screening force ${\bf F}_C$ resulting from other electrons as they are redistributed in the inclusion by the external field, and the electromotive force ${\bf F}_L$ due to currents produced by electrons as they move under the action of the field. Here, $m$ and $e<0$ are electron mass and charge, ${\bf v}$ is its velocity, and $\gamma$ is the decay frequency for the metal (a friction coefficient for electrons). By determining the dynamics of the electrons in the metal and by averaging this dynamics over the meta-atom, its polarizability can be determined, and one can introduce effective dielectric parameters of the entire metamaterial.

It should be noted that averaging and homogenization are two distinct procedures. From the averaging of the microscopic parameters we get the average displacement of electrons and the overall polarizability of the metal ring segments. Specifically, the averaging yields the resonance frequency and spectral shape. Once the polarizability of the metal segments is known, the effective medium parameters can be introduced by homogenization. To be able to connect the properties of PCMs with those of bulk planar chiral media, we employ the standard approaches used for bulk metamaterials. However, it should be realized that the resulting effective parameters can be attributed to the PCMs under study only in specific cases (for the normal incidence of light).

The screening force ${\bf F}_C$ is expected to depend heavily on the geometry of the metallic inclusion. Under the action of the field, all the electrons are displaced, and uncompensated charges are accumulated at the edges of the metal in the direction of the field. If the distance between the edges in that direction is small and the edges are wide enough, the induced charges produce the field similar to that inside a capacitor, i.e. it is mostly homogeneous and compensates the external field. Therefore it can be assumed that the electrons simply won't move between such edges. So, for our geometry we can neglect  out-of-plane and radial electron motion as long as $d$ and $h$ are small.

%%%%%%%%%%%%%%%%%%%%%%%%%%%
\begin{figure}[tb]
\centerline{\includegraphics[width=0.98\columnwidth]{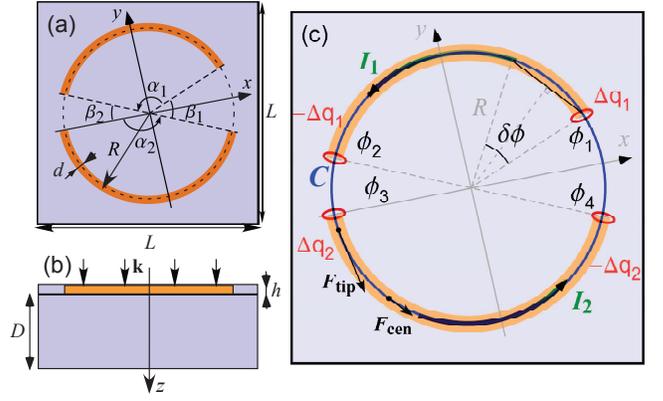}}
\caption{(Color online) (a) Top view and (b) side view of a CSR unit cell;
(c) schematics of the induced charge redistribution in a CSR:
$\Delta q_{1,2} = \Sigma_{1,2} \Delta S$, where $\Delta S = h d$.
The parameters are as in Ref.~\onlinecite{PlumCSR} (cell
size: $L=15$mm, ring radius: $R=6$ mm, ring width: $d=0.8$ mm,
ring thickness: $h=35$ $\mu$m, substrate thickness: $D=1.6$ mm, substrate material: dielectric with $\varepsilon_d=4$). The angles determining the CSR composition are $\alpha_1=140^\circ$, $\alpha_2=160^\circ$, $\beta_1=40^\circ$, $\beta_2=20^\circ$.
%Blue (light) and red (dark) colors denote the dielectric substrate and the metallic inclusions, respectively.
}
\label{fig:1}
\end{figure}
%%%%%%%%%%%%%%%%%%%%%%%%%%%

The displacement of an individual electron $\zeta$ therefore depends on time and on the position in the ring: $\zeta=\zeta(t,\varphi)$. It is subject to the equation of azimuthal motion for the  electrons ($\dot\zeta\equiv\partial\zeta/\partial t$):
\begin{equation}
m \ddot{\zeta} + m \gamma \dot{\zeta} =  \eee_\varphi \cdot
(e \ee + \mathbf{F}_C + \mathbf{F}_L).
\end{equation}

Equation of motion introduces resonant features in the response of the structure. Electrons are pushed by the external driving forces, which induce both electric currents in the ring and charges at the tips of the ring segments. The charges create a ``restoring force'' $F_C$ like the one in a harmonic oscillator (e.g., by Hooke's law). The electromotive force $F_L$ caused by the current is the reason of the additional non-dynamic inertia of the electrons. As in the ordinary classical mechanics, a particle under these forces will be subject to resonant motion, which should manifest itself as resonant properties of the effective medium.

Usually, the electron displacement is small and in good accordance with the external field, so for a subwavelength-sized inclusion it can be considered as position-independent. In this case, $\zeta(t,\varphi)=s(t)$. However, this behavior is expected to break down in the vicinity of a particle plasmon resonance where the electron shift can be substantial and more in accordance with the resonant mode. Therefore, $\zeta$ becomes very sensitive to the $\varphi$-dependence.
At the fundamental resonance, it is reasonable to assume the standing-wave positional dependence in the two ring segments
\begin{equation}
\begin{gathered}
\zeta_1(t,\varphi) = s_1(t) (\pi/2) \sin(\pi(\varphi-\varphi_1)/\alpha_1), \\
\zeta_2(t,\varphi) = s_2(t) (\pi/2) \sin(\pi(\varphi-\varphi_3)/\alpha_2),
\end{gathered}\label{eq:zeta_phi}\end{equation}
where  $s_{1} = (1/\alpha_{1})\int_{\varphi_1}^{\varphi_2} \zeta_{1} d\varphi$ and
$s_{2} = (1/\alpha_{2})\int_{\varphi_3}^{\varphi_4} \zeta_{2} d\varphi$
have the meaning of an averaged electron displacement in the corresponding segment.
It is seen that the electrons do not move near the tips of the split
rings as $\zeta(t,\varphi_i) = 0$, $i = 1,\ldots,4$ (see Fig. \ref{fig:1}), and
maximally shift in the middle of the segments. While the positional dependence in $\zeta$
can be safely neglected in the off-resonant situations,
it will be shown below that it is required to correctly predict the resonant frequency of a CSR.

The effective response of the entire meta-atom is obtained by averaging over all available azimuthal angles where metal is present, i.e.,  for $\varphi$ from
$\varphi_1$ to $\varphi_2$ and from $\varphi_3$ to $\varphi_4$ [see Fig.~\ref{fig:1}(c)]. With the brackets $\langle A \rangle$ denoting the averaging of $A$ over  $\varphi$, the equation of motion for the averaged electron displacement $s$ takes the form
\begin{equation}
m \ddot{s} + m \gamma \dot{s} = e \langle (\eee_\varphi \cdot
\ee^d) \rangle + \langle \eee_\varphi \cdot\mathbf{F}_C \rangle
+\langle \eee_\varphi \cdot\mathbf{F}_L \rangle. \label{eq_mot}
\end{equation}
We assume that the metal is fully embedded into the substrate medium, so external field $\ee$ coincides with the field in the ambient medium $\ee^d$.

\subsection{Screening force and induced charges}

Since the accumulation of charges along the ring edges only serves to prevent the
electron motion in the radial direction, the screening force $\mathbf{F}_C$ inside the ring segment
can be approximately described as the Coulomb force generated by
the charges at the tips of the segment, labeled by the points $\varphi_1$ and $\varphi_2$ ($\varphi_3$ and $\varphi_4$).
This assumption (which is valid outside the immediate vicinity of the tips) is further
substantiated by the fact that charges do accumulate mostly near the tips of metallic objects.
In a CSR, we need to take into account the tips of the other segment in the same split ring [see Fig. \ref{fig:1}(c)], as well as the influence of the neighboring CSRs.

The screening force on electrons in a ring segment from its own tips is essentially similar to
that in a metallic rod with width $d$ and thickness $h$. For the rod placed in dielectric with $\varepsilon=\varepsilon_d$, the force is equal to
\begin{equation}
F^\text{tip}(x) = -\frac{4 e \Sigma}{\varepsilon_d} \arctan\left( \frac{h d/4}{x \sqrt{x^2 + h^2/4 + d^2/4}} \right), \label{force_rod}
\end{equation}
where $\Sigma$ is the surface charge density at the tip facet of the rod with area $\Delta S = d\times h$ , and
$x$ is the distance from the center of the facet to an observation point.

To avoid unphysical behavior of the force $F^{\text{tip}}$ very close to the tip, the positional dependence of the displacement $\zeta(\varphi)$ in Eq.~\eqref{eq:zeta_phi} has to be taken into account as $\Sigma = e N \zeta(t,\varphi)$.
%The resulting force near the tip will account for the extra $\sin$-dependence due to $\zeta$.

The expression given by Eq.~\eqref{force_rod} is approximately valid if $x$ is small enough so that the arc segment with length $x$ is not significantly different in shape from a rod with dimensions $x\times d \times h$. It can be assumed so if the line connecting the observation point with the tip lies wholly within the metal, i.e., if $0<x<2\sqrt{Rd}$ for $h\ll d$.

Outside of that range, Eq.~\eqref{force_rod} no longer holds. However, for the observation point far away from the tips one can regard the charges accumulated at the tip as one point charge and calculate the screening force according to the Coulomb law. Unlike $F^{\text{tip}}$, the forces ``in the center'' of an arc segment (labeled $F^\text{cen}$) are not dominated by contribution from any particular tip, so both tips of the segment in question, as well as the tips of the neighboring segment in the ring and  from the neighboring rings need to be taken into account. As a result, we get the following expression for the upper segment

\begin{eqnarray}
 F_1^\text{cen}& =& {\bf F}_C \cdot \eee_\varphi
 \nonumber \\ &=&
 \frac{e \Sigma_1 \Delta S}{\varepsilon_d}
\sum_{n,m} \left(\frac{\eee_\varphi(\varphi) ({\bf L}_{nm} + R \eee_r(\varphi_1))}
{|R \eee_r(\varphi) - {\bf L}_{nm} - R \eee_r(\varphi_1)|^3}
 \right. \nonumber \\ && \left.
-\frac{\eee_\varphi(\varphi) ({\bf L}_{nm} + R \eee_r(\varphi_2))}
{|R \eee_r(\varphi) - {\bf L}_{nm} - R \eee_r(\varphi_2)|^3} \right) \nonumber \\
& +& \frac{e \Sigma_2 \Delta S}{\varepsilon_d}
\sum_{n,m} \left(\frac{\eee_\varphi(\varphi) ({\bf L}_{nm} + R \eee_r(\varphi_3))}
{|R \eee_r(\varphi) - {\bf L}_{nm} - R \eee_r(\varphi_3)|^3}
 \right. \nonumber \\ && \left.
- \frac{\eee_\varphi(\varphi) ({\bf L}_{nm} + R \eee_r(\varphi_4))}
{|R \eee_r(\varphi) - {\bf L}_{nm} - R \eee_r(\varphi_4)|^3} \right),
\label{force_cen} \end{eqnarray}
where ${\bf L}_{nm} = n \eee_x + m \eee_y$ is the position of a CSR in the ($x$, $y$) plane,
characterized by a couple of lattice indices $n$ and $m$.
The charge densities for the upper and lower segments are $\Sigma_1 = e N \zeta_1(t,\varphi)$
and $\Sigma_2 = e N \zeta_2(t,\varphi)$, respectively.

To obtain the averaged screening force for the meta-atom, we assume that it is given by Eq.~\eqref{force_rod} for $0<x<2\sqrt{Rd}$ (i.e., within an angle $\delta \varphi = 2 \sqrt{d/R}$ from each tip, see Fig.~\ref{fig:1}(c)), and by Eq.~\eqref{force_cen} elsewhere. Accordingly, the averaging over the upper segment results in the formula
\begin{eqnarray}
\langle \eee_\varphi {\bf F}_C \rangle_1 = \frac{1}{\alpha_1}\left(
\int_{\varphi_1}^{\varphi_1 + \delta \varphi} F^\text{tip} d\varphi +  \int_{\varphi_1 +
\delta \varphi}^{\varphi_2 - \delta \varphi} F^\text{cen}_1 d\varphi + \right. \nonumber \\
\left. \int_{\varphi_2 - \delta \varphi}^{\varphi_2} F^\text{tip} d\varphi \right),
\label{eq:screening}
\end{eqnarray}
Similar expressions can be obtained for the lower segment. These integrals cannot be evaluated in closed form, but can be easily calculated numerically.  The resulting screening force for the two segments in the ring can be finally expressed as
\begin{equation}
F_{1,2} = - k_{11,22} s_{1,2} - k_{12,21} s_{2,1}, \label{Fscr}
\end{equation}
where $k_{ij}$ are coefficients obtained from the integrals in Eq.~\eqref{eq:screening}. They have the meaning of "stiffness" coefficients in a mechanical oscillator, and are given by

\begin{widetext}
\begin{eqnarray}
&& k_{11,22} =  \frac{m \omega_p^2}{\alpha_{1,2} R \varepsilon_d} \int_{0}^{R \delta\varphi} \sin\left( \frac{\pi x}{\alpha_{1,2} R} \right) \arctan\left( \frac{h d/4}{x \sqrt{x^2 + h^2/4 + d^2/4}} \right) dx \nonumber \\
&& - \frac{m \omega_p^2 h d}{8 \alpha_{1,2} \varepsilon_d} \int_{\varphi_{1,3} + \delta\varphi}^{\varphi_{2,4} - \delta\varphi} \sin\left( \frac{\pi (\varphi - \varphi_{1,3})}{\alpha_{1,2}} \right)
\sum_{n,m} \left[\frac{\eee_\varphi(\varphi) ({\bf L}_{nm} + R \eee_r(\varphi_{1,3}))} {|R \eee_r(\varphi) - {\bf L}_{nm} - R \eee_r(\varphi_{1,3})|^3}
-\frac{\eee_\varphi(\varphi) ({\bf L}_{nm} + R \eee_r(\varphi_{2,4}))}
{|R \eee_r(\varphi) - {\bf L}_{nm} - R \eee_r(\varphi_{2,4})|^3} \right] d\varphi,  \\
&& k_{12,21} = - \frac{m \omega_p^2 h d}{8 \alpha_{1,2} \varepsilon_d} \int_{\varphi_{1,3} + \delta\varphi}^{\varphi_{2,4} - \delta\varphi} \sin\left( \frac{\pi (\varphi - \varphi_{2,1})}{\alpha_{2,1}} \right)
\sum_{n,m} \left[\frac{\eee_\varphi(\varphi) ({\bf L}_{nm} + R \eee_r(\varphi_{3,1}))}
{|R \eee_r(\varphi) - {\bf L}_{nm} - R \eee_r(\varphi_{3,1})|^3}
- \frac{\eee_\varphi(\varphi) ({\bf L}_{nm} + R \eee_r(\varphi_{4,2}))}
{|R \eee_r(\varphi) - {\bf L}_{nm} - R \eee_r(\varphi_{4,2})|^3} \right] d\varphi. \nonumber
\end{eqnarray}
\end{widetext}

\subsection{Electromotive force}

The force ${\bf F}_L = e {\bf E}_{em}$ appears because the electron motion along the split ring under the action of the external electric field can be regarded as currents in the loop and induces secondary magnetic field, which in turn penetrates the closed contour of the loop
and creates an electromotive force. This force is directed in opposition
to the induced current in accordance with Lenz's rule.

To determine ${\bf F}_L$, we start from the Maxwell equation in integral form
\begin{equation}
\int_C \ee_{em} d{\bf l} = -\frac{1}{c} \dot{B}_z S.
\end{equation}
The integration is performed along the loop of the ring $C$, while
$S=\pi R^2$ is its area and $B_z$ is the $z$-component of magnetic field.
Assuming homogeneous electromotive electric field over the contour and taking the length of
the loop as $2 \pi R$ and its area as $S = \pi R^2$, the force affecting
an electron in a CSR is
\begin{equation}
\langle \eee_\varphi \cdot\mathbf{F}_L \rangle = -\frac{e R}{2 c} \dot{B}_z.
\end{equation}
Since the magnetic field of a normally incident wave has no $z$-component,
$B_z$ can only originate from the current flowing along the loop.
The field can be estimated from the Biot-Savart law
\begin{equation}
{\bf B}(\rr_0) = \int \frac{I d{\bf l}\times \rr}{c r^3},
\label{eq:induction}
\end{equation}
where $\rr$ is the radius-vector from the current element $I d{\bf l}$
to the observation point $\rr_0$. The current $I$ in
the metallic parts of the loop is $I_{1,2} = e N \dot{\zeta}_{1,2} \Delta
S = e N \dot{\zeta}_{1,2} h d$.

We then assign the magnetic field in the loop to be the magnetic field at the center
of the split ring. The magnetic field created by the current in each segment is summed:
\begin{equation}
B_z = \frac{1}{c R} \left( I_1 \alpha_1 + I_2 \alpha_2 \right),
\end{equation}
so the electromotive force finally equals
\begin{eqnarray}
F_L = \langle \eee_\varphi \cdot\mathbf{F}_L \rangle = - m \frac{\omega_p^2 \Delta S}{8\pi c^2} \left[ \alpha_1 \ddot{s}_1
+ \alpha_2 \ddot{s}_2 \right]. \label{Fem}
\end{eqnarray}

The force $F_L$ is proportional to the electron's acceleration which allows us to regard the coefficient in front of $\ddot{s}_{1,2}$ as an effective mass.

\subsection{Equation of motion}

By substituting the screening Eq. (\ref{Fscr}) and electromotive Eq. (\ref{Fem}) forces into equation of
motion for the electrons (\ref{eq_mot}) in each segment,
we derive the following coupled differential equations:
\begin{eqnarray}
m_{11} \ddot{s}_{1} + m_{12} \ddot{s}_2 + \gamma \dot{s}_{1} + k_{11} s_1 + k_{12} s_2 = e (\langle\eee_\varphi\rangle_{1} \ee^d), \nonumber \\
m_{21} \ddot{s}_{1} + m_{22} \ddot{s}_2 + \gamma \dot{s}_{2} + k_{22} s_2 + k_{21} s_1 = e (\langle\eee_\varphi\rangle_{2} \ee^d),\label{deq}
\end{eqnarray}
where
\begin{eqnarray}
m_{11} &=& m + m \frac{\alpha_1 \omega_p^2 h d}{8\pi c^2}, \qquad
m_{12} = m \frac{\alpha_2 \omega_p^2 h d}{8\pi c^2}, \nonumber \\
m_{21} &=& m \frac{\alpha_1 \omega_p^2 h d}{8\pi c^2}, \qquad
m_{22} = m + m \frac{\alpha_2 \omega_p^2 h d}{8\pi c^2}.
\label{deq:coefs}
\end{eqnarray}
Here
$\langle\eee_\varphi\rangle_{1} = \frac{1}{\varphi_2-\varphi_1}
\int_{\varphi_1}^{\varphi_2} \eee_\varphi d\varphi$ and $
\langle\eee_\varphi\rangle_{2} = \frac{1}{\varphi_4-\varphi_3}
\int_{\varphi_3}^{\varphi_4} \eee_\varphi d\varphi$ denote the
averaging of the vector $\eee_\varphi$ over the upper and lower
segment, respectively (see Fig.~\ref{fig:1}).

For the geometrical parameters of CSRs used, it can be estimated that   $m \alpha_1 \omega_p^2 h d/8\pi c^2 \gg m$,
which leads to $m_{11} \approx m_{21}$ and $m_{12} \approx m_{22}$. In other words, the effective
mass of an electron is dominated by the contribution of the electromotive force
rather than by the dynamic counterpart. However, electromotive and dynamic masses can become comparable at the nanoscale.
%For greater sizes of an element than nano-, we can certainly use the electromotive mass instead of the dynamic one.

Note that the effective mass $m_{ij}$ can be regarded as inductance, while
the coefficients $k_{ij}$ have the meaning of inverse capacitance. In this picture, it is seen that the proposed model coincides with a well-known effective-circuit (LC) model with the effective inductance and capacitance calculated from the material and geometrical properties of the CSR. This result is not surprising because the LC model is expected to be valid for millimeter-sized resonators at the gigahertz operating frequencies. It also follows that  the LC-model
remains applicable as long as $\Delta S = h d \gg 8\pi c^2/(\alpha_1 \omega_p^2) \sim 10^{-13}$ m$^2$.

In a CSR, the coupling between the two segments is realized by means of the effective masses
$m_{12}$ and $m_{21}$ and the ``stiffness'' coefficients
$k_{12}$ and $k_{21}$. As we have just seen, all mass coefficients $m_{ij}$ are almost identical. It is not the case for the coefficients $k_{ii}$, which contain the contribution in the vicinity of the tips $F^\text{tip}$ [see Eq.~\eqref{eq:screening}]. This contribution is greater than the forces in the central part of the ring $F^\text{cen}$ by about 10 times, so it can be considered dominant. On the contrary, the coefficients $k_{ij\neq i}$ contain only $F^\text{cen}$, so $k_{ij} \ll k_{ii}$. Moreover, since $k_{ii} \simeq m_{ii} \omega^2 \approx m_{ij} \omega^2 \gg k_{ij}$, so the condition $F^\text{cen}\ll F^\text{tip}$ allows the contribution of $k_{ij\neq i}$ to be entirely neglected.
Thus, we can simplify the expressions for the remaining stiffness coefficients $k_{ii}$:
\begin{eqnarray}
k_{11,22} = \frac{m \omega_p^2}{\varepsilon_d \alpha_{1,2} R} \int_0^{R \delta\varphi} \sin(\pi x/(\alpha_{1,2} R))\nonumber\\
\times\arctan\left( \frac{h d/4}{x \sqrt{x^2 + h^2/4 + d^2/4}} \right) dx. \label{k1122}
\end{eqnarray}
Note that both tips for each segment are taken into account.

Eqs.~(\ref{deq}) are essentially equations of motion for two coupled oscillators with a
time-harmonic driving force. Therefore, we look for a solution of these equations (\ref{deq})
in the form $s_{1,2} = \exp(-\rmi \omega t) l_{1,2}$,
where $\omega$ is the frequency of the
incident wave and $l_{1}$ and $l_2$ are constants.
Then the equations are simplified as
\begin{equation}\begin{gathered}
(- m_{11} \omega^2 - \rmi \omega \gamma + k_{11}) l_{1} +(- m_{12} \omega^2 + k_{12}) l_2 = e \langle\eee_\varphi\rangle_{1} \ee^d, \\
(- m_{21} \omega^2 + k_{21})l_{1} + (- m_{22} \omega^2 - \rmi \omega \gamma + k_{22}) l_2 = e \langle\eee_\varphi\rangle_{2} \ee^d.
\end{gathered} \label{system}\end{equation}

This system is easily solved with respect to  $l_1$
and $l_2$. So, we find the averaged displacements of electrons in both CSR segments of
the ring:
\begin{eqnarray}
l_1 = \frac{1}{4\pi N e} (\chi_{11}(\omega) \langle\eee_\varphi\rangle_{1} +
\chi_{12}(\omega) \langle\eee_\varphi\rangle_{2}) \ee^d , \nonumber
\\
l_2 = \frac{1}{4\pi N e} (\chi_{21}(\omega) \langle\eee_\varphi\rangle_{1} +
\chi_{22}(\omega) \langle\eee_\varphi\rangle_{2}) \ee^d,
\label{avel}
\end{eqnarray}
where
\begin{eqnarray}
&& \chi_{11}(\omega) = \frac{-m\omega_p^2(m_{22} \omega^2 + \rmi \gamma \omega
- k_{22})}{D_0}, \nonumber \\
&& \chi_{12}(\omega) = \frac{m\omega_p^2 (m_{12} \omega^2 - k_{12})}{D_0}, \nonumber \\
&& \chi_{21}(\omega) = \frac{m\omega_p^2 (m_{21} \omega^2 - k_{21})}{D_0}, \nonumber \\
&& \chi_{22}(\omega) = \frac{-m\omega_p^2(m_{11} \omega^2 + \rmi \gamma
\omega - k_{11})}{D_0}, \nonumber \\
&& D_0 = (m_{11} \omega^2 + \rmi \gamma \omega - k_{11})(m_{22} \omega^2 +
\rmi \gamma \omega - k_{22}) \nonumber \\
&& - (m_{12} \omega^2 - k_{12}) (m_{21} \omega^2 - k_{21}).
\label{chi}
\end{eqnarray}

\subsection{Polarizability of the meta-atom}

Finally, to arrive at the effective polarization of the unit metamaterial cell, we
present it as a sum of the polarization of both CSR segments and of the surrounding dielectric:
\begin{eqnarray}
4 \pi \langle {\bf P}\rangle &=& (1-p_1-p_2) (\varepsilon_d -1) \ee^d  \nonumber \\
&+& 4\pi N e \left( p_1 \langle \eee_\varphi \rangle_1 s_1
+ p_2 \langle \eee_\varphi \rangle_2 s_2 \right), \label{polariz0}
\end{eqnarray}
where $p_{1,2}={R d \alpha_{1,2} h}/{L^2 D}$ are filling factors of the two metallic
segments. By substituting the average displacement (\ref{avel})
into Eq. (\ref{polariz0}), we get
\begin{eqnarray}
4 \pi \langle {\bf P} \rangle &=& (1-p_1-p_2) (\varepsilon_d -1) \ee^d \nonumber \\
&+& [p_1 \langle \eee_\varphi \rangle_1 \otimes
(\chi_{11} \langle\eee_\varphi\rangle_{1} +
\chi_{12} \langle\eee_\varphi\rangle_{2}) \nonumber \\
&+& p_2 \langle \eee_\varphi \rangle_2 \otimes (\chi_{21}
\langle\eee_\varphi\rangle_{1} + \chi_{22}
\langle\eee_\varphi\rangle_{2})] \ee^d,
\end{eqnarray}
where $\langle\ee^m\rangle$ is the thickness-averaged electric field in metal, $\ee^d$ is the field in dielectric, and ${\bf a} \otimes {\bf b}$ denotes an outer (tensor or dyadic) product between two vectors.

%Electric field in dielectric $\ee^d$ in terms of the incident field $\ee$
%can be estimated from the Fresnel formulae:
%\begin{equation}
%\ee^d = \frac{2}{\sqrt{\varepsilon_d}+1} \ee.
%\end{equation}
The average electric field in the metamaterial cell approximately equals
the electric field in dielectric, $\langle \ee \rangle \approx \ee^d$.
So, the polarization of the cell can be expressed as
\begin{eqnarray}
&& 4 \pi \langle {\bf P} \rangle = 4 \pi \chi \langle \ee \rangle = \left[ (\varepsilon_d -1) \right. \nonumber \\
&& \left. +  \tilde{\chi}_{11}(\omega) \langle
\eee_\varphi \rangle_1 \otimes \langle\eee_\varphi\rangle_{1} +
\tilde{\chi}_{12}(\omega) (\langle \eee_\varphi \rangle_1 \otimes
\langle\eee_\varphi\rangle_{2} \right. \nonumber \\
&& \left. + \langle \eee_\varphi \rangle_2
\otimes \langle \eee_\varphi\rangle_{1}) + \tilde{\chi}_{22}(\omega)
\langle \eee_\varphi \rangle_2 \otimes
\langle\eee_\varphi\rangle_{2} \right] \langle \ee \rangle,
\end{eqnarray}
where
\begin{equation}
\tilde{\chi}_{ij} = p_i \chi_{ij} \qquad (i, j = 1, 2).
\end{equation}
Notice that $\tilde{\chi}_{12} = \tilde{\chi}_{21}$, so the
susceptibility tensor $\chi$ is symmetric, as would be required by the
reciprocity considerations. \cite{Tretyakov}

\section{Effective parameters of a PCM\label{sec:3}}

\subsection{Permittivity and permeability tensors\label{sec:3a}}

To move on from a unit cell to the entire lattice of meta-atoms comprising a PCM, one needs to take into account that the field in each unit cell is modified by the presence of the neighboring meta-atoms. (It is known that an array of symmetric meta-atoms with no intrinsic chirality can exhibit {\em extrinsic} chiral properties due to the way the atoms are arranged in a lattice. \cite{KseniaPRA09})
If the arrangement of meta-atoms is not too dense so that the individual atoms remain distinct, it can be assumed that the influence of the neighboring atoms is weak and can be simulated by regarding the meta-atoms as effective dipoles. The resulting field in
each cell is equal to the sum of the electric field averaged over the whole planar metamaterial $\overline{\ee}$ and the field of the dipoles: $\langle \ee \rangle = \overline{\ee} + 4 \pi
\hat{C} \langle {\bf P} \rangle$, where $\hat{C}$ is the interaction
matrix. \cite{Ishimaru} The interaction matrix depends on the symmetry of the lattice. For a planar arrangement of atoms (see Appendix A) its form is
\begin{equation}
\hat{C} = 0.36 \frac{D}{\varepsilon_d L} \left({\bf e}_x
\otimes {\bf e}_x + {\bf e}_y \otimes {\bf e}_y - 2 {\bf e}_z
\otimes {\bf e}_z \right).
\label{eq:intmatrix}\end{equation}
 For the case $D\ll L$, the interaction matrix
has negligible components, and the average field in the metamaterial nearly coincides with that in a single meta-atom: $\langle\ee\rangle \approx \overline{\ee}$. In our case $D/L \sim 0.1$, therefore, the influence of the surrounding meta-atoms can be non-negligible.

%%%%%%%%%%%%%%%%%%%%%%%%%%%

\begin{figure}[tb]
%\centerline{ \includegraphics[width=1.0\columnwidth,clip]{ReEps.eps}
%\includegraphics[width=1.0\columnwidth,clip]{ImEps.eps}}
\centerline{ \includegraphics[width=1.0\columnwidth,clip]{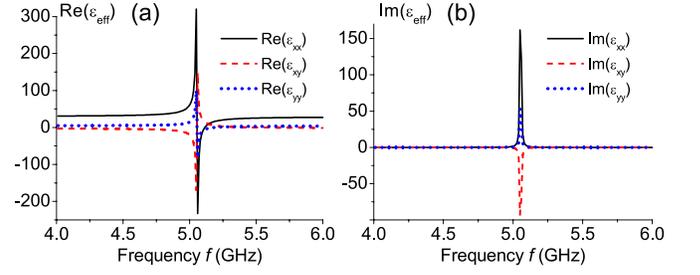}}
\caption{(Color online) (a) Real and (b) imaginary parts of the components of the
effective dielectric permittivity tensor for the split-ring
metamaterial with geometrical parameters given in the caption of
Fig. \ref{fig:1}. In calculations, we neglect the force $F^\text{cen}$,
so that $k_{12} = k_{21} = 0$ and $k_{11}$ and $k_{22}$ are given by
Eq. (\ref{k1122}). The copper ring is characterized by $\omega_p =
2000$ THz, $\gamma=8$ THz; permittivity of the dielectric is
$\varepsilon_d = 4$.}
\label{fig:eps}
\end{figure}

%%%%%%%%%%%%%%%%%%%%%%%%%%%

The resulting effective dielectric permittivity tensor of the PCM can be derived from equations $\langle \ee \rangle = \overline{\ee} + 4 \pi \hat{C} \langle {\bf P} \rangle$, $\varepsilon_\text{eff} \overline{\ee} = \overline{\ee} + 4 \pi \langle {\bf P} \rangle$, and $\langle {\bf P} \rangle = \chi \langle \ee \rangle$. Its final form is
\begin{eqnarray}
&& \varepsilon_\text{eff}(\omega) = 1 + 4 \pi (\chi^{-1} - 4 \pi \hat{C})^{-1}
\label{eps_eff}
\end{eqnarray}
and can be rewritten as a matrix:
\begin{equation}
\varepsilon_\text{eff} = \left( \begin{array}{ccc} \varepsilon_{xx} &
\varepsilon_{xy} & 0 \\ \varepsilon_{xy} & \varepsilon_{yy} & 0 \\
0 & 0 & \varepsilon_{zz} \end{array} \right). \label{eps_matrix}
\end{equation}
The components $\varepsilon_{ij}$ are complex, and it is seen that $\varepsilon_\text{eff}$ has the structure of a dichroic and anisotropic medium, in line with crystallographic expectations. \cite{ourOL}

The typical frequency dependencies of $\varepsilon_{xx}$, $\varepsilon_{yy}$, and $\varepsilon_{xy}$ for the example CSR structure in Fig.~\ref{fig:1} are shown in Fig.~\ref{fig:eps}. It can be seen that the structure features a Lorentz-like absorption resonance in the range near 5.1~GHz, in agreement with experimental results for such CSRs.\cite{PlumCSR} This resonance corresponds to the minimum in the denominator $D_0$ in Eq.~\eqref{chi}, so it is an intrinsic excitation in an individual meta-atom.
%In this example and in the further investigation we will neglect the force acting on electrons in the central part of the ring. This approximation will be discussed in connection with the transmission spectra in Section \ref{sec:4}.

It is important to realize two fundamental limitations of the presented homogenization approach. First, we neglect the magnetic dipole and electric quadrupole contributions (they have the same order of magnitude and should be accounted for simultaneously \cite{Falk08}). This can be safely done for the light normally incident onto a planar structure. Indeed, the quadrupole moment $Q$ has the form $Q(z) = (Q_\perp + Q_{zz} \eee_z \otimes \eee_z) \exp(\rmi k_0 z)$, where $Q_\perp \eee_z = \eee_z Q_\perp = 0$ and the $z$-dependence in $Q(z)$ is caused by the external field. Hence the quadrupole contribution
%to the electric induction
has the form $\nabla Q(z) = \rmi k_0 \eee_z Q(z)$ and is $z$-directed, resulting in a contribution to $\varepsilon_{zz}$. The magnetic dipole moment is $z$-directed, too, since the electrons move in the ($x$, $y$)-plane, which results in effective magnetic permeability $\mu = {\rm diag}(1,1,\mu_{zz})$. Therefore, these higher-moment contributions will not affect $\varepsilon_{xx}$, $\varepsilon_{yy}$, and $\varepsilon_{xy}$, which are the only components that will play a part in determining the normal-incidence transmission and reflection spectra. So it is sufficient to consider just the electric dipole moment, assuming $\mu = 1$ for the PCM.

Secondly, and perhaps more seriously, the presented approach is commonly employed for bulk metamaterial homogenization, \cite{Baena} and its applicability for metamaterial surfaces leaves room for uncertainty regarding how, specifically, the homogenization in the $z$-direction should be performed. It is questionable whether a planar surface can be described as a finite-thickness slab of a bulk effective medium that would mimic the response of the metamaterial for all cases of the incident light, even when magnetic dipole and electric quadrupole contributions are taken into account. It is commonly assumed that first-principle characterization methods based on  multipole expansion \cite{Falk09,Falk08,Alu} should be used instead of homogenization. Still, we can obtain the effective material parameters valid in a specific case, in order to see whether a bulk material with planar chiral properties  can be related to real PCMs.\cite{ourOL} Hence we can continue with the effective permittivity derived in Eq.~\eqref{eps_matrix}, keeping in mind that it is only valid for normal incidence of light.

\subsection{Polarization eigenstates}

Polarization of the eigenwaves of anisotropic medium with
dielectric permittivity (\ref{eps_matrix}) can be found from the
wave equation\cite{FedorovBook}
\begin{equation}
\left( n^2 (1-\eee_z \otimes \eee_z) - \varepsilon_\text{eff} \right) \ee =0,
\end{equation}
where $n$ is the effective refractive index for the eigenwave. The electric
fields in two eigenstates take the form
\begin{equation}
\ee_\pm = E_x \left( 1, \frac{\varepsilon_{xx}-\varepsilon_{yy}\pm
\sqrt{(\varepsilon_{xx}-\varepsilon_{yy})^2+4\varepsilon_{xy}^2}}{2
\varepsilon_{xy}}, 0 \right).
\end{equation}

\begin{figure}[bt]
\centerline{ \includegraphics[width=0.8\columnwidth,clip]{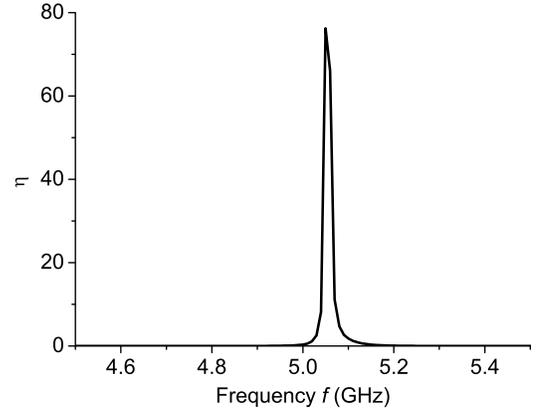}} \caption{
Coefficient $\eta$ vs. frequency calculated for parameters indicated
in Fig.~\ref{fig:eps}.}
\label{fig:eta}
\end{figure}

Since the components of permittivity tensor are complex, these eigenwaves are elliptically polarized. Their direction of rotation (``right- or left-handedness'')
can be defined by the parameter $\eta_\pm = (|\varepsilon_{xy}/E_x|)^2 \rmi \eee_z (\ee_\pm
\times \ee_\pm^\ast)$.  \cite{FedorovBook} An RH-polarized wave has $\eta_\pm>0$, an LH-polarized one has $\eta_\pm<0$, and, obviously, $\eta_\pm=0$ coresponds to a linearly polarized wave whose sense of handedness cannot be determined.

Consequently, we can identify whether the calculated effective parameters of the CSR metamaterial correspond to those of a PCM by simply evaluating the product  $\eta = \eta_+ \eta_-$. Conventionally, in isotropic or lossless birefringent media $\eta=0$ as the eigenwaves are linearly polarized. In 3D chiral or Faraday media, $\eta<0$ as the eigenwaves (either circularly or elliptically polarized, depending on the presence of anisotropy in addition to optical activity) are counter-rotating. On the contrary, PCMs (and elliptically dichroic crystals, see Ref.~\onlinecite{ourOL}) are characterized by co-rotating polarization eigenstates, so it is expected that $\eta>0$ in these media.

Fig.~\ref{fig:eta} shows the coefficient $\eta$ calculated for the CSR metamaterial in Fig.~\ref{fig:1} with dielectric permittivity shown in Fig.~\ref{fig:eps}. Indeed, it can be seen that $\eta$ is positive for all frequencies in the vicinity of the intrinsic resonance around 5.1 GHz. Thus, the signature crystallographic property of a PCM (co-rotating elliptical polarization eigenstates) is indeed reproduced in the effective medium, confirming that it is 2D rather than 3D chirality that manifests in CSR metamaterials.

\subsection{Transmission and reflection spectra}

As the final step in the theoretical model, we briefly outline the calculation procedure for the optical spectra of a PCM. Following the set-up in Ref.~\onlinecite{PlumCSR}, we consider the CSR metamaterial of effective thickness $D+h\approx D$ and calculate the reflection and transmission coefficients of a dichroic and anisotropic monolayer with dielectric permittivity given by Eq.~(\ref{eps_eff}). Since we are interested in all possible polarizations of the incident wave, it is convenient to make use of the well known covariant operator generalization of the transfer matrix method (the covariant Fedorov's approach \cite{FedorovBook}). For the details on this method, the reader is referred to previous publications. \cite{BBL,BorzdovTheory,ourJOA}

We define a unit vector ${\bf q}=\mathbf{e}_z$ pointing in the propagation direction and write the Maxwell equations for a monochromatic normally incident wave in the form
\begin{equation}
{\bf q}^\times \frac{{\rm d}}{{\rm d} z} \hh = - \rmi k_0
\varepsilon \ee, \qquad {\bf q}^\times \frac{{\rm d}}{{\rm d} z} \ee
= \rmi k_0 \hh,
\label{eq:maxwell}
\end{equation}
where $k_0=\omega/c$ is the vacuum wavenumber,
% $\omega$ is the wave circular frequency,
and ${\bf q}^\times$ denotes the antisymmetric
tensor dual to the vector ${\bf q}$ [$({\bf
q}^\times)_{ik}=\mathcal{E}_{ijk} q_j$, $\mathcal{E}_{ijk}$ is the Levi-Civita pseudotensor]. \cite{BBL}

The fields are always tangential, and the field vectors are continuous across the layer interfaces. Eqs.~\eqref{eq:maxwell} can be combined into the form
\begin{equation}
\frac{{\rm d} \ww(z)}{{\rm d} z}=\rmi k_0 M \ww(z),
\label{Meqt_planar}
\end{equation}
where, for non-magnetic, non-gyrotropic  materials,
\begin{equation}
\ww = \left(\begin{array}{c}\hh \\
{\bf q} \times \ee\end{array}\right),
\qquad
M=\left(\begin{array}{cc}0&-{\bf q}^\times\varepsilon {\bf q}^\times \\
I&0\end{array}\right).
\label{ABCD_planar}
\end{equation}
Here $I=1-{\bf q} \otimes {\bf q}= - {\bf q}^{\times 2}$ is the
projection operator onto the plane normal to ${\bf q}$, and 1 is
the three-dimensional identity tensor.
%Since the fields ${\bf q}\cdot \ee={\bf q}\cdot \hh=0$, only two components of the field vectors need to be retained. Hence the block matrix $M$, which contains all the information about the material properties, has dimensions $4\times 4$.
The fundamental solution of Eq. (\ref{Meqt_planar}) is a matrix
exponential
\begin{equation}
\ww(z) = P(z) \ww(0), \qquad P(z) = \exp(\rmi k_0 M z),
\label{eq:evolution}
\end{equation}
where $4 \times 1$ dimensional constant vector $\ww(0)$ is the
initial field. The $4\times 4$ matrix $P(z)$ is called the evolution
operator.

Taking into account that in the medium surrounding the metamaterial layer (i.e., in air) the fields are related as $ {\bf q}\times \ee=\pm I \hh$ depending on the propagation direction,\cite{BorzdovTheory} the incident and reflected waves at the input (air/PCM) boundary are related to the initial field $W(0)$ as
\begin{equation}
\ww(0) = \left( \begin{array}{c} I \\ I \end{array} \right)
\hh_\text{inc} + \left( \begin{array}{c} I \\ -I
\end{array} \right) \hh_\text{refl}.
\end{equation}
From Eq.~\eqref{eq:evolution}, the field at the
%PCM/substrate interface is $P_{m}(c_0) \ww(0)$ and at the
output (PCM/air) interface is  $P(D) \ww(0)$. The evolution
operator of the metamaterial $P$ can be
derived by setting $\varepsilon=\varepsilon_\text{eff}$ in Eq.~(\ref{ABCD_planar}).

On the other hand, the field at
the output interface is the transmitted wave
\begin{equation}
\ww(D) = \left( \begin{array}{c} I \\ I
\end{array} \right) \hh_\text{tr}.
\end{equation}
Hence the boundary conditions take the form
\begin{equation}
\left( \begin{array}{c} I \\ I \end{array} \right) \hh_\text{tr} = P
\left[ \left( \begin{array}{c} I \\ I
\end{array} \right) \hh_\text{inc} + \left( \begin{array}{c} I \\ -I
\end{array} \right) \hh_\text{refl} \right]. \label{BC}
\end{equation}
%Eq.~(\ref{BC}) is the system of algebraic equations to be solved for the unknown magnetic fields $\hh_{tr}$ and $\hh_{refl}$.
Multiplying Eq.~(\ref{BC}) by the rectangular block matrix
$\left( \begin{array}{cc} I & I \end{array} \right) P^{-1}$
and thus eliminating $\hh_\text{refl}$, the expression for the transmitted magnetic field becomes
\begin{equation}
\hh_\text{tr} = 2 \left[ \left( \begin{array}{cc} I & I \end{array}
\right) P^{-1} \left( \begin{array}{c} I \\ I \end{array} \right)
\right]^{-1} \hh_\text{inc},
\label{transm_field}
\end{equation}
which, along with the evolution operator $P$, will depend on $\omega$. Finally we define the transmission coefficient of the metamaterial slab as the ratio between the intensity of
transmitted and incident waves:
\begin{equation}
T(\omega) = \frac{|\hh_\text{tr}(\omega)|^2}{|\hh_\text{inc}|^2}. \label{transmission}
\end{equation}
Eqs.~(\ref{transm_field}) and (\ref{transmission}) hold likewise for the electric fields. \cite{BorzdovPart}

\section{Comparison with numerical simulations\label{sec:comp}}

\begin{figure*}[tb]
\centerline{ \includegraphics[width=0.8\textwidth]{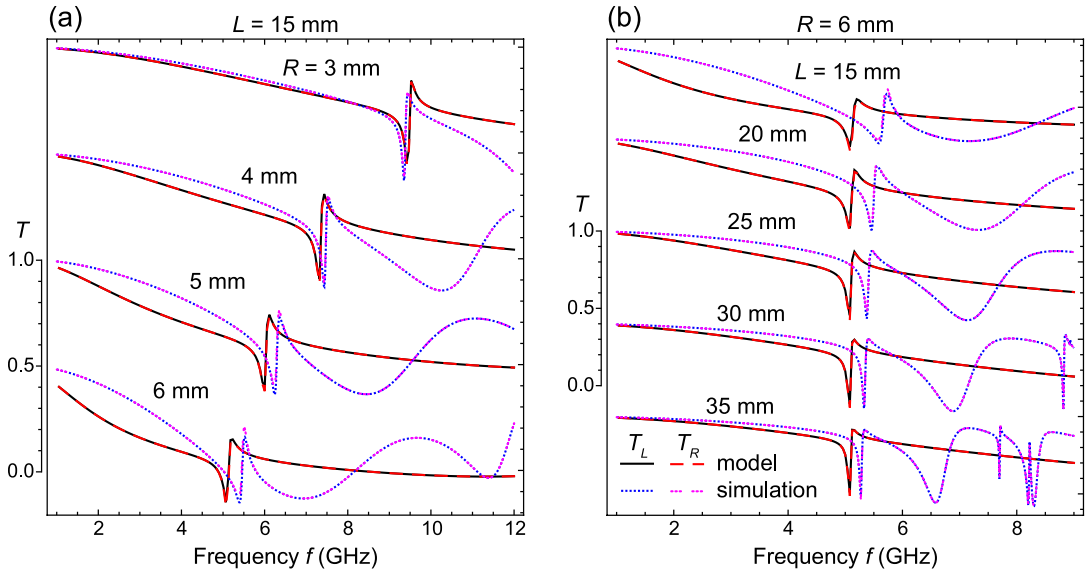}}
\caption{(Color online) Comparison between the proposed model and the numerical simulation (CST Microwave Studio) for the transmission spectra for left-handed ($T_L$)  and right-handed ($T_R$) circularly polarized incident waves. Calculations are done for asymmetric ring metamaterial with $\alpha_1 = 140^\circ$, $\alpha_2 = 160^\circ$,
$\beta_1 = 40^\circ$, and $\beta_2 = 20^\circ$ for different ring radiuses and lattice constants: (a) for $L=15$ mm and varying $R$, (b) for $R=6$ mm and varying $L$. Other  parameters are as given in Figs. \ref{fig:1} and \ref{fig:eps}. }
\label{fig:spectra}
\end{figure*}

\begin{figure*}[tb]
\centerline{ \includegraphics[width=0.8\textwidth]{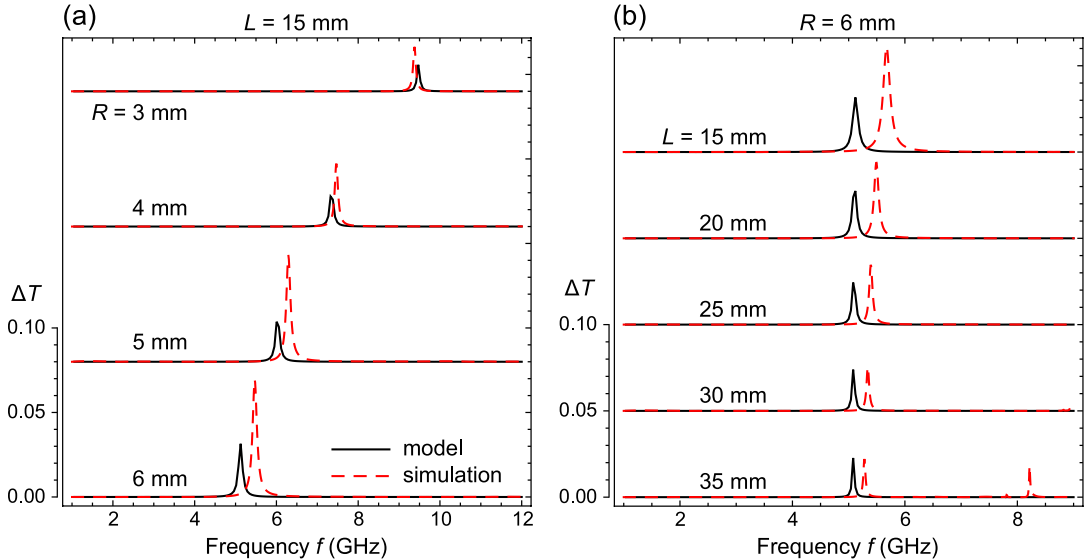}}
\caption{Same as Fig.~\ref{fig:spectra} but for the transmission difference of $\Delta T=T_L-T_R$. }
\label{fig:spectraDT}
\end{figure*}

Varying the frequency and polarization of the incident wave $\hh_\text{inc}$ , one can obtain the corresponding transmission spectrum as outlined in the previous section. We will be particularly interested in investigating the PCM transmittance for LH vs.~RH circularly polarized incident wave, labelled $T_L(\omega)$ and $T_R(\omega)$, and corresponding to complex vectors $\hh_\text{inc}= \frac{1}{\sqrt{2}} (\eee_x \pm \rmi \eee_y)$, respectively. Here, $T_L(\omega)$ and $T_R(\omega)$ are the overall transmittances, without regard for polarization of transmitted light. The vast majority of materials (either naturally occurring or artificial) do not discriminate between LH and RH circular polarization in transmittance, so that $\Delta T(\omega)=T_L(\omega)-T_R(\omega)$ is zero for all frequencies. A non-zero $\Delta T$ signifies the presence of circular dichroism and enantiomeric asymmetry.

To test the applicability limits of the proposed model, we first compare analytically and numerically calculated $T_{L}(\omega)$ and $T_{R}(\omega)$ for our example CSR structure of Fig.~\ref{fig:1} with varying ring radius $R$ and lattice period $L$. Numerical results are obtained using a commercially available finite integration solver (CST Microwave Studio) in the frequency domain, using periodic boundary conditions in the $x-y$ directions.

We see that a resonant dip in the transmission that results from the intrinsic resonance for the components of $\varepsilon$ in the analytical model (see Fig.~\ref{fig:eps}) is reproduced in numerical calculations and corresponds to the fundamental dipole excitation of the CSR. The resonant frequency $f_\text{res}$ changes when the ring radius is varied [Fig.~\ref{fig:spectra}(a)]. The resonance has a Fano-like shape, which is also reproduced numerically.

For frequencies below $f_\text{res}$, we see a good agreement between analytical and numerical results, which gradually worsens as $R$ is increased in comparison with $L$, so that meta-atoms become closer to each other and the assumptions about a sparsely packed lattice that were needed in deriving Eq.~\eqref{eps_eff} become increasingly violated. This also causes a mismatch between analytically and numerically derived $f_\text{res}$. On the other hand, for frequencies above $f_\text{res}$ the agreement is worse because the numerical spectra are affected by higher-order CSR resonances (which are explicitly not accounted for in our determination of a CSR's response), as well as the Bragg resonances of the lattice, which are also neglected in our account under the assumption that the effective medium is regarded as homogeneous.

Fig.~\ref{fig:spectra}(b) shows the dependence of the analytical vs.~numerical spectra as $L$ is varied for the constant CSR radius. It is seen that the agreement below $f_\text{res}$ is restored as $L$ increases, confirming our reasoning. However, the agreement above $f_\text{res}$ becomes much worse because the Bragg resonances scale as  $f_\text{Bragg} \sim 1/L$, and are thus pushed into lower frequencies. Physically it means that the upper frequency limit where $L \ll \lambda$ (and where the structure can be regarded as a metamaterial) becomes smaller. Ultimately $f_\text{Bragg}$ moves past $f_\text{res}$, which is where the response of an individual meta-atom becomes irrelevant to the whole optical properties of the structure.

Therefore as far as the quantitative agreement of $T(\omega)$ is concerned, the model is found to be valid for the lattice period $L$ not exceeding 20-25 mm and for the CSR radius $R$ significantly lower than $L$. Most discrepancies occur in higher frequencies ($f>f_\text{res}$) and can be attributed to higher-order and/or lattice resonances which have been left out of consideration intentionally. The trade-off between split-ring coupling and grating diffraction is mentioned in the recent Ref.~\onlinecite{Sersic}.

Good agreement within these validity limits of the model can also be confirmed in the spectra for the asymmetry $\Delta T(\omega)=T_{L}-T_{R}$   (Fig.~\ref{fig:spectraDT}). Moreover, we see that in a vast majority of cases, $T_{L}\approx T_{R}$ everywhere except the vicinity of $f_\text{res}$, which coincides with the range where polarization eigenstates are co-rotated elliptical (see Fig.~\ref{fig:eta}). So, many quantitative discrepancies in the transmission spectra have no effect over $\Delta T$ and the model remains qualitatively valid for all the parameter values shown in Fig.~\ref{fig:spectraDT} with mismatch to the resonance frequency $f_\text{res}$ and the maximum value of $\Delta T(f_\text{res})$ gradually increasing as the approximations behind the presented model become less accurate. The exception is the case of larger $L$ where non-zero $\Delta T$ is also seen at odd-numbered higher-order CSR resonances [Fig.~\ref{fig:spectraDT}(b)]. However, since such additional chiral response is spectrally well separated from the fundamental resonance that interests us, it does not affect planar chiral properties of CSRs under present investigation.

Hence, numerical simulations confirm that the proposed microscopic description of the CSR structures reproduces the PCM behavior, as reported in previous experiments. \cite{PlumCSR,ZhelPRL} Within the assumptions of the model that takes into account only the fundamental particle plasmon resonance of the CSR segments, the model provides a good agreement in a wide range of parameters.  Having established this, we move on to investigate how the spectrum $\Delta T (\omega)$ behaves in various CSR designs.

\section{Geometrical transformations with chiral split rings\label{sec:geom}}
%\section{Spectral properties of chiral split-ring PCM with varying geometry\label{sec:4}}

A non-zero $\Delta T$, signifying the presence of circular dichroism, carries special significance for 2D structures. One notices that spatial inversion of the whole system with respect to the plane normal to the metamaterial changes the handedness of the circular polarization (LH $\leftrightarrow$ RH), {\em and} replaces the structure with its enantiomeric counterpart (see Fig.~\ref{fig:PCMs}). Hence, if $T(\omega)$ is the spectrum of any planar structure and $\tilde{T}(\omega)$ is the spectrum of its enantiomeric counterpart, then
\begin{equation}
T_{L,R}(\omega)=\tilde{T}_{R,L}(\omega)
\label{eq:enant}
\end{equation}
for any structure at any frequency. Therefore, for any planar meta-atom with no distinct 2D enantiomers (i.e., with an in-plane symmetry axis) $\Delta T=0$. So, it is important to point out that $\Delta T \neq 0$ indicates the presence of planar chirality.	
A CSR of the design considered here (see Fig.~\ref{fig:1}) becomes symmetric and therefore achiral if either the ring segments are of equal length ($\alpha_1=\alpha_2$), or the gaps between the segments are equal ($\beta_1=\beta_2$), or else in a few degenerate cases when there is effectively just one segment (i.e., $\alpha_1=0$, $\alpha_2=0$, $\beta_1=0$, or $\beta_2=0$).

Furthermore, the reversal of the direction of incidence also transforms the structure into its enantiomeric counterpart but does {\em not} change the incident wave polarization. Hence, if the transmission spectra for the forward- vs.~backward-incident wave are labeled $T^{f,b}$, respectively, then Eq.~\eqref{eq:enant} results in
\begin{equation}
T^{f,b}_{L,R}(\omega)=\tilde{T}^{f,b}_{R,L}(\omega)=T^{b,f}_{R,L}(\omega),
\label{eq:direct}
\end{equation}
again, for any structure at any frequency.\cite{Tretyakov}
Therefore, a non-zero $\Delta T$ is a measure of the planar structure's directional asymmetry and its magnitude can be used as an indication of how strongly the planar chiral properties of a structure manifest themselves optically.

%\subsection{Geometrical transformation of chiral split rings}

\begin{figure}[tb]
\centerline{ \includegraphics[width=1.0\columnwidth,clip]{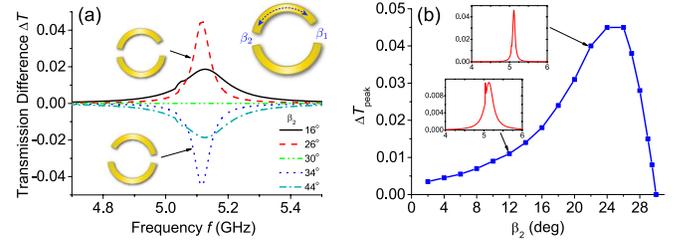}}
\caption{(Color online) Difference of transmission for LH and RH circularly polarized incident waves for different ring geometries with $L=15$ mm, $R=6$ mm,  $\alpha_1 = 140^\circ$, $\alpha_2 = 160^\circ$, $\beta_1 = 60^\circ-\beta_2$: 
 (a) comparison of $\Delta T$ for enantiomeric counterparts ($\beta_1 \leftrightarrow \beta_2$) in the range $10^\circ \leq \beta_2 \leq 50^\circ$;
(b) evolution of maximum $\Delta T$ ($\Delta T_\text{peak}$) between symmetric and asymmetric CSRs for $\beta_2\leq30^\circ$.}
\label{fig:a1a2}
\end{figure}

Consider first the displacement of one of the ring segments along the circle by varying $\beta_1$  and setting $\beta_2=360^\circ - \beta_1-(\alpha_1+\alpha_2)$. The results are presented in Fig.~\ref{fig:a1a2}. Not surprisingly, chiral properties are rather weak for $\beta_1\gg\beta_2$ and become larger as $\beta_2$ increases towards the case of Fig.~\ref{fig:spectra} where $\beta_1 = 40^\circ$, $\beta_2 = 20^\circ$. After a certain optimum value, however, $\Delta T$ decreases again, vanishing completely in the symmetric case $\beta_1 = \beta_2 = 30^\circ$. So the proposed theory confirms that both circular dichroism and directional asymmetry indeed vanish when mirror symmetry is present.

The spectral shape $\Delta T (\omega)$ is seen to have a wider shape for small $\beta_2$, becoming the narrowest for the optimal case and then diminishing without significantly changing its shape. This is what one would expect as the inter-segment coupling (which is stronger for smaller $\beta_2$ because the tips are in close proximity) pushes the particle plasmon resonances of a CSR apart from each other. This feature is specific to CSR design: while the response of each arc-shaped segment in a CSR closely resembles that of a rod of equal length,\cite{ourFiO10} the split ring is a geometry where the tips of the segments are in much closer proximity than for the rods placed at similar distance. Hence, the field enhancement near the tips causes the response of the whole CSR to depend strongly both on the individual segments and on inter-segment coupling.

In addition, making $\beta_2$ small while maintaining the CSR orientation should increase the contribution of extrinsic effects to chiral properties. \cite{KseniaPRA09} This is indeed seen in Fig.~\ref{fig:a1a2}(b), and this is likely the reason of a small but non-vanishing $\Delta T$ for these values.

Displacing the ring segment past the symmetric case $\beta_1 = \beta_2 = 30^\circ$, we notice in Fig.~\ref{fig:a1a2}(a) that $\Delta T$ changes its sign and that an exchange of $\beta_1 \leftrightarrow \beta_2$ results in the inversion of planar chiral properties ($\Delta T(\omega)\leftrightarrow - \Delta T(\omega)$). This entirely confirms the result expected from Eq.~\eqref{eq:enant}, taking into account that structures obtained by an exchange of $\beta_1$ and $\beta_2$ are 2D enantiomeric counterparts.

\begin{figure}[bt]
%\centerline{ \includegraphics[width=0.7\columnwidth,clip=]{a1b2.eps}}
%\centerline{\includegraphics[width=0.7\columnwidth,clip=]{a2b2.eps}}
\centerline{\includegraphics[width=1.0\columnwidth,clip=]{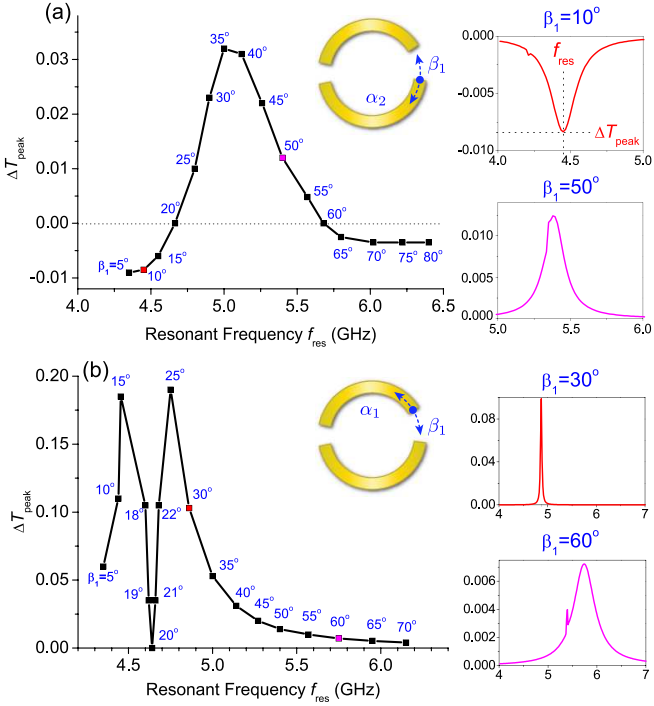}}
 \caption{(Color online)
Difference of transmission for LH and RH circularly polarized incident waves: (a) for $\alpha_1 = 140^\circ$, $\beta_2 = 20^\circ$, varying $\beta_1$ and
$\alpha_2=200^\circ-\beta_1$; (b) for $\alpha_2 = 160^\circ$, $\beta_2 = 20^\circ$, varying $\beta_1$ and $\alpha_1=180^\circ-\beta_1$.}
\label{fig:a1b2}
\end{figure}

We also examine CSRs with variable length of the longer ring segment by varying  $\beta_1$ and $\alpha_2$ while keeping the other two angles constant (note that all the time $\alpha_1+\alpha_2+\beta_1+\beta_2=360^\circ$). The results are given in Fig.~\ref{fig:a1b2}(a). As expected, $\Delta T$ vanishes for the two symmetric cases $\beta_1 = \beta_2 =20^\circ$ and
$\beta_1 = 60^\circ$ ($\alpha_1 = \alpha_2 = 140^\circ$). The sign of $\Delta T$ changes when these two symmetric cases are traversed.
%Once again we see the transition from the narrow-peak $\Delta T$ for smaller $\beta_1$ to a wider-peak shape that was observed in Fig.~\ref{fig:a1a2}.
The increase of the peak frequency $f_\text{res}$ corresponds to an increase in the stiffness coefficients $k_{11,12}$ as one of the segments becomes shorter, affected by a change in the coupling between the segments' resonances as they differ in length more strongly.
%For $\beta_1>60^\circ$, a redistribution of magnitude in $\Delta T$ is observed, so that the peaks are lower but wider. This can also be influenced by extrinsic effects because making one of the segments significantly shorter than the other makes the meta-atom less circle-like, and the PCM becomes geometrically closer to a tilted-cross array. \cite{KseniaPRA09}
%Note that the bandwidth of circular dichroism is increased in this regime, which can be useful in designing broadband PCMs.

Similarly varying the length of the shorter ring segment, i.e., changing the angles $\beta_1$ and $\alpha_1$, the same behavior is observed, as can be seen in Fig.~\ref{fig:a1b2}(b). The spectral shape changes in the same manner as in the previous case, the peak in $\Delta T$  becoming broader for larger $\beta_1$. However, this case is specific because $\alpha_2+\beta_2=180^
\circ$, so there is only one symmetric shape ($\beta_1=\beta_2=20^\circ$, $\alpha_1=\alpha_2=160^\circ$). Because of this ``degeneracy'', $\Delta T$ does not change sign when traversing the symmetric case. The same property can be responsible for higher peak values of $\Delta T(f_\text{res})$.

\begin{comment}
In both cases, one can note the appearance of a secondary narrow resonance in $\Delta T$
when the segments have significantly different length \sergei{Near $60^\circ$ in Fig 7(a), the lengths of the segments are close, but the peaks can be distinguished. At the same time, only at $40^\circ$ the second peak arises in Fig. 7(b). I am not sure that we can just say that the lengths should be different}. This resonance has a Fano-type line shape and can be attributed to the interaction between a narrow resonance from ... and a wide resonance from ... . \andrey{Please complete this sentence briefly.} \sergei{I do not know exactly, how these resonances appear. We have a resonance for permittivity at, let say, frequency $f_0$. In transmission spectrum, we can have 2 resonances. Broad resonance at a frequency not equal to $f_0$ and narrow resonance approximately at $f_0$. Generally we can say that "This resonance has a Fano-type line shape and can be attributed to the interaction between a narrow resonance from antisymmetric excitation and a wide resonance from in-phase current oscillations."  }
\end{comment}

%%% \subsection{Scaling and coupling effects}

\begin{figure}[tb]
\centerline{ \includegraphics[width=1.0\columnwidth,clip=]{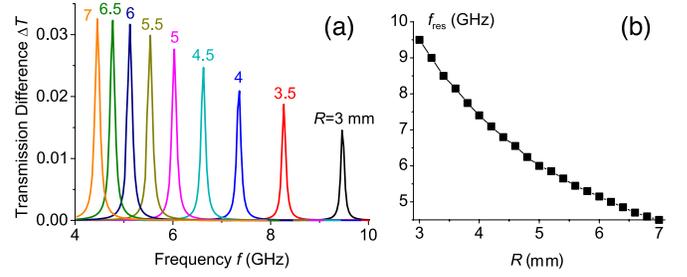} }
\caption{(Color online) (a) Difference of transmissions of LCP and RCP incident waves
for different values of ring radius $R$.
%The rest geometrical parameters are the same as in Fig. \ref{fig:1}.
(b) Dependence of the peak frequency $f_\text{res}(R)$.}
 \label{fig:R}
\end{figure}

Finally, to determine how the chiral properties of PCMs scale with the size of the meta-atoms, we investigate the dependence of $\Delta T$ on the ring radius $R$ in Fig.~\ref{fig:R}. The resonant frequency $f_\text{res}$ depends on the radius primarily due to $1/R$ appearing in the stiffness coefficients $k_{ij}$ [see Eq.~\eqref{k1122}]. The estimated dependence $f_\text{res} \sim 1/R$ is confirmed in Fig.~\ref{fig:R}(b). The resonance peak in $\Delta T (\omega)$ broadens as $R$ increases.

\section{Concluding remarks\label{sec:5}}

In conclusion, we have proposed a microscopic theoretical description of planar chiral metamaterials based on the electronic Lorentz theory. Using a chiral split-ring (CSR) geometry\cite{PlumCSR} as an example and considering the dynamics of individual free electrons, we arrive at expressions for electromagnetic response of a single split-ring meta-atom.
Its polarizability is derived analytically without phenomenological parameters.
The effective dielectric permittivity tensor $\varepsilon_\text{eff}$ is then obtained from the single-atom response
along the lines of standard homogenization techniques, \cite{Ishimaru,Vinogradov} to the extent that these techniques can be used at all to describe planar structures. \cite{Yatsenko,Falcone,Holloway}
%\andrey{Please insert references}
Even in this simplified consideration, the crystallographic structure of this tensor is shown to coincide with that of elliptically dichroic media, as would be expected from earlier theoretical studies.\cite{ourOL}

The transmission spectra of a CSR-based PCM are then calculated using the standard wave operator based extension of transfer-matrix methods. \cite{BBL,BorzdovTheory} In agreement with previous theoretical and experimental results, \cite{ZhelPRL,PlumCSR} the spectra show a difference with respect to whether the incident wave has left- or right-handed circular polarization [$\Delta T(\omega)=T_L(\omega)-T_R(\omega)$]. This difference, which translates to directional asymmetry, \cite{PlumCSR} is shown to strongly depend on the geometrical parameters of the CSR (Figs.~\ref{fig:a1a2}--\ref{fig:a1b2}). Whenever the split ring is symmetric, i.e., when there are no distinguishable 2D enantiomers, optical manifestation of planar chirality is seen to vanish [$\Delta T(\omega)=0$], and $\Delta T$ changes its sign when the structure is replaced with its enantiomeric counterpart [Fig.~\ref{fig:a1a2}(a)].
%Optical manifestation of planar chirality is found to be the strongest when the structure has the least degree of in-plane mirror symmetry (Fig.~\ref{fig:a1a2}).
%, and is markedly diminished by direct inter-segment coupling (Fig.~\ref{fig:coupl}).The results obtained can be used to optimize and tailor the performance of such PCMs.

Note that we have deliberately chosen the overall CSR orientation in the lattice so as to focus on purely intrinsic chirality and to suppress extrinsic effects where possible. A detailed investigation of how intrinsic and extrinsic chiral properties interact in CSR-based PCMs
%(which can already be seen in some cases in  Figs.~\ref{fig:a1a2}--\ref{fig:a1b2})
warrants a separate investigation.

%This suggests that mixing several types of meta-atoms in a 2D lattice would enhance the asymmetric transmission of PCMs via extrinsic effects.

While the specific split-ring geometry is chosen for its relative simplicity in analytical derivation, it should be understood that the proposed approach can be extended to any planar meta-atom consisting of thin wire-like metallic elements where transverse motion of electrons is restricted. The integrals in Eqs.~\eqref{eq:screening} and \eqref{eq:induction} are likely to be more complicated and may have to be taken numerically. Moreover, it may be particularly challenging to determine the correct charge density dependencies [see Eq.~\eqref{eq:zeta_phi}] and to identify the loops that contribute to the effective mass; one may need to use equivalent LC-circuit or even resort to using data from direct numerical simulations if the geometry is particularly intricate. Nevertheless, once the equations for the effective mass and stiffness coefficients for a particular unit cell geometry are established, parametric transformations of this geometry lend themselves to very easy semi-analytical treatment within the proposed framework.

\begin{comment}
\sergei{General algorithm (can be translated to the main body of the manuscript).
(i) Define a unit cell.
(ii) Using symmetry of the structure, analyze which currents and charges can appear. Think whether the system is resonant or not (this is needed to make an assumption about correct dependence of electron's displacement).
(iii) Currents forming a closed loop introduce magnetic field which, in its turn, creates electromotive force. Find this force $F_L$. If the system has no closed currents (e.g. 1D grating), calculation of $F_L$ can be omitted.
(iv) Taking into account the influence of surrounding cells (if necessary), write the screening force $F_C$.
(v) By substituting the accounted forces $F_L$ and $F_C$ into equation of motion, solve this equation and find the displacements of the electrons. Write down polarization of the unit cell.
(vi) Using interaction matrix, take into consideration the fields of all unit cells. Find the effective parameters of metamaterial.
(vii) For a given thickness of the metamaterial slab, calculate transmission/reflection spectra.
}\end{comment}

Furthermore, oblique wave incidence and non-planar meta-atoms composed of similar thin elements are also tractable if the magnetic dipole and electric quadrupole contributions are accounted for, giving rise to corrections in $\varepsilon_\text{eff}$ and introducing effective magnetic permeability tensor $\mu$, as well as gyration pseudotensors responsible for magnetoelectric coupling or spatial dispersion.
\cite{BBL,BorzdovPart,BorzdovTheory} 
However, it still remains an open question whether such oblique-incidence treatment would be universal taking into account the inherent limitations of applying volume homogenization to surface structures. 
\cite{Yatsenko,Falcone,Holloway}
%\andrey{Insert references here too, please.}
Should such a generalization prove feasible, it is very interesting to extend the proposed approach from a single-layer PCM to PCM-based multilayers (in the cases when such PCMs can lend themselves to 3D homogenization \cite{Andriyeuski}) and investigate its applicability as a planar metamaterial turns into a bulk one.

\begin{acknowledgments}
Inspiring discussions with D. N. Chigrin, V. Fedotov, and A. Chipouline are acknowledged. This work was supported in part by
the Danish Research Council for Technology and Production Sciences (THz COW),
Basic Research Foundation of Belarus (F10M-021), the
Deutsche Fouschungsgemeinschaft (DFG Research Unit FOR 557), and the 
Natural Sciences and Engineering Research Council of Canada (NSERC).
\end{acknowledgments}
\bigskip

\appendix
\section{Derivation of the interaction matrix\label{appendix:a}}

To calculate the interaction matrix for meta-atoms arranged in a square planar array used in Sec.~\ref{sec:3a}, one can write the field in an $n$th unit cell as a sum of the averaged electric field in the
metamaterial $\overline{\ee}$ and the fields
of the dipoles ${\bf p}$ at the center of the each cell:
\begin{equation}
\langle \ee \rangle=\overline{\ee} + \sum_{i \neq n} \frac{3 {\bf r}_i \otimes
{\bf r}_i - r_i^2}{\varepsilon_d r_i^5} {\bf p},
\end{equation}
where ${\bf r}_i$ is the radius-vector of $i$th dipole. Following Ref.~\onlinecite{Ishimaru}, we define
the interaction matrix $\hat{C}$ by means of equation
$\langle \ee \rangle=\overline{\ee} + 4 \pi \hat{C} {\bf P}_0$. The polarization of the
medium is connected with the dipole moment of the single cell via
${\bf P}_0 = N_0 {\bf p}$, where $N_0 = 1/(L^2 D)$ is the number of inclusions per unit
volume (we suppose that the cell is square). Therefore, the interaction matrix equals
\begin{equation}
\hat{C} = \frac{L^2 D}{4 \pi \varepsilon_d} \sum_{i \neq n}\frac{3 {\bf r}_i
\otimes {\bf r}_i - r_i^2}{r_i^5}.
\end{equation}

Placing the origin at the center of the cell under consideration
($n$th cell) we present the radius-vectors of the others as ${\bf
r}_{i}={\bf r}_{k l}= L (k {\bf e}_{x} + l {\bf e}_{y})$, where
integer numbers $k$, $l$ vary from $-\infty$ to $\infty$. The
$n$th cell is characterized by the numbers $k=0$ and $l=0$ and
should be excluded from the summing.

The result of the summation  is used as Eq.~\eqref{eq:intmatrix} in Sec.~\ref{sec:3}:
\begin{equation}
\hat{C} = \frac{\tau D}{\varepsilon_d L} \left( I - 2 {\bf e}_z \otimes {\bf
e}_z \right), \label{int_mat_is}
\end{equation}
where $I = {\bf e}_x \otimes {\bf e}_x + {\bf e}_y \otimes {\bf
e}_y$ is the projection operator onto the plane normal to the $z$ axis,
\begin{equation}
\tau = \frac{1}{2 \pi} \left( \zeta(3) +
\sum_{k,l=1}^\infty \frac{1}{(k^2+l^2)^{3/2}} \right) \approx 0.3592
\end{equation}
and $\zeta(x) = \sum_{k=1}^\infty k^{-x}$ is the Riemann Zeta
function.

%%% ====================================================================
%%% ====================================================================
%%% ====================================================================

\end{document}